\documentclass[aps,prb,superscriptaddress,showpacs,twocolumn]{revtex4-1}
\usepackage{graphicx}

\begin{document}
%
%
\title{Vibrational density of states of silicon nanoparticles}

\author{R. Meyer} \email{rmeyer@cs.laurentian.ca}
\affiliation{Department of Mathematics and Computer Science,
Laurentian University, 935 Ramsey Lake Road, Sudbury (Ontario) P3E
2C6, Canada}

\author{D. Comtesse} \affiliation{Fachbereich Physik,Universit\"{a}t
Duisburg-Essen, Lotharstra\ss e 1, D-47048 Duisburg, Germany}

%
%
\begin{abstract} The vibrational density of states of silicon
nanoparticles in the range from 2.3 to 10.3\,nm is studied with the
help of molecular-dynamics simulations. From these simulations the
vibrational density of states and frequencies of bulk-like vibrational
modes at high-symmetry points of the Brillouin-zone have been
derived. The results show an increase of the density of states at low
frequencies and a transfer of modes from the high-frequency end of the
spectrum to the intermediate range. At the same time the peak of
transverse optical modes is shifted to higher frequencies. These
observations are in line with previous simulation studies of metallic
nanoparticles and they provide an explanation for a previously
observed discrepancy between experimental and theoretical data
[C. Meier \textit{et al.}, Physica E, \textbf{32}, 155 (2006)].
\end{abstract} 
\pacs{63.22.Kn,62.25.Jk,81.05.Cy,02.70.Ns}
\date{November 19, 2010}
\maketitle

%
%
\section{Introduction} Modern nanotechnology is leading to the
development of a new generation of applications and devices based on
nanostructured materials. An important part of the development of many
nano-devices is the thermal design. If the heat generated inside a
device cannot be transported away efficiently, the system will fail or
even break due to overheating. Thermal properties like heat capacity
and thermal conductivity as well as many other material properties are
strongly influenced by the vibrational density of states (VDOS). For
this reason a profound understanding of the laws governing the
vibrational properties of nanostructured materials is of high
technological and fundamental interest.

This article focuses on the vibrational properties of silicon
nanoparticles. Silicon is an important engineering material whose bulk
properties have been studied extensively. Its favourable electrical
properties have made it the base material for all microelectronics and
advanced technology exists for the manufacturing of nanostructured
silicon based devices.

The vibrational properties of bulk crystalline Si have been studied
both experimentally using neutron scattering
techniques\cite{Dolling:63a,Nilsson:72a} as well as theoretically
using empirical and first-principles
methods.\cite{Pandey:73a,Giannozzi:91a,Wei:94a} In addition to this,
Raman scattering experiments have been used in numerous studies of the
vibrational properties of silicon
nanostructures.\cite{Richter:81a,Furukawa:88a,Sui:92a,Munder:92a,%
Fujii:96a,Tanino:96a,Wang:99a,Saviot:04a,Adu:05a,Meier:06a,Ristic:09a}
In many of these studies it is found that the first order peak due to
the transverse optical (TO) phonon at the $\Gamma$ point is broadened
and its maximum shifted to lower energies. The phonon-confinement
model\cite{Richter:81a,Campbell:86a} attributes this behaviour to the
fact that in nanostructures more vibrational modes can become Raman
active than in bulk crystals since the translational symmetry is
broken and consequently no $\Delta\mathbf{k}=0$ selection rule
applies. In Ref.~\onlinecite{Meier:06a}, Meier et al. observed
pronounced decreases of the Raman shift in Silicon nanoparticles. A
comparison with calculated values based on the phonon-confinement
model showed the experimentally observed decrease of the Raman shift
to be considerably stronger than predicted by the model
calculations.\cite{Meier:06a}

Theoretically, the vibrational properties of nanostructured silicon
have also been studied. Heino has calculated phonon dispersion
relations for silicon thin films using molecular-dynamics
simulations\cite{Heino:07a} whereas Saviot \textit{et al.} have
studied low frequency vibrational modes in silicon nanoparticles using
an elastic continuum model.\cite{Saviot:04a} Hu \textit{et al.} and
Valentin \textit{et al.} employed empirical atomistic force models to
calculate the VDOS of silicon nanoparticles by diagonalization of the
dynamical-matrix.\cite{Hu:01a,Valentin:07a} The latter two studies did
not include relaxation effects from the particle surfaces.

In this work, we use molecular-dynamics simulations in combination
with the modified embedded-Atom method
(MEAM)\cite{Baskes:89a,Baskes:92a} potential for silicon from
Ref. [\onlinecite{Baskes:94a}] to calculate the vibrational density of
states and selected zone-boundary phonon frequencies for silicon
nanoparticles with diameters in the range 2.3 -- 10.3\,nm. These
simulations include surface effects and the results show changes from
the bulk density of states that are similar to those observed in
metallic nanoparticles.\cite{Kara:98a,Meyer:02a,Meyer:03a,Meyer:07a}
While the limitations of the potential do not allow us to make
quantitative predictions, our calculations show clear trends in the
developement of the particle VDOS as the particle size is reduced. In
particular, we observe an increase of the VDOS at low frequencies, a
shift of the high frequency modes to higher frequencies and a transfer
of weight from the TO mode peak to intermediate frequencies. Our
results provide an explanation for the discrepancy observed in
Ref.~\onlinecite{Meier:06a}.

%
%
\section{\label{SecComp}Computational Details} The calculations of
vibrational properties of Si nanoparticles presented in this work are
all based on molecular-dynamics simulations using the MEAM potential
for Si from Ref.~\onlinecite{Baskes:94a}. The advantage of the MEAM
formalism\cite{Baskes:89a,Baskes:92a} is that it is not only suitable
for simulations of pure Si in the diamond structure but that it has
also been used successfully in studies of composite systems combining
Si with other elements like, Ni,\cite{Baskes:94a} Mb,\cite{Baskes:99a}
or Au.\cite{Kuo:04a} A disadvantage of the potential is that it
severely overestimates the phonon frequencies in
silicon.\cite{Heino:07a} Qualitatively the potential gives however a
fair description of silicon and reflects the characteristic features
of its phonon dispersion relations. Since in this work we are more
interested in the qualitative changes of the VDOS than quantitatively
exact frequencies and since we hope to exploit the potential's
transferability in future work focusing on composite systems, we
decided to use this potential despite its shortcomings.

%
%
\begin{table}
\caption{Approximate diameters $d$ of the model particles and the
number of atoms $N$ they contain.}
\begin{tabular}{c|cccccc}%
\hline\hline%
$d \mathrm{(nm)}$ & 2.3 & 2.8 & 4.1 & 6.1 & 8.2 & 10.3 \\%
\hline%
$N$ & 329 & 657 & 1803 & 5815 &14175 & 27943\\%
\hline\hline
\end{tabular}
\label{TabSize}
\end{table} In order to study the influence of particle size on the
vibrational properties in Si nanoparticles, we constructed a set of
model particles by cutting a spherical region out of an ideal diamond
lattice. The diameters of the particles and the number of atoms they
contain are summarized in Table~\ref{TabSize}. In addition to the
nanoparticle configurations we used a bulk-like configuration
containing 1,000,000 Si atoms arranged on an ideal $50\times 50\times
50$ cells diamond lattice.

Prior to the calculation of vibrational frequencies, we carefully
equilibrated our model systems at $T=300\,\mathrm{K}$. In order to
determine the equilibrium lattice constant of the bulk configuration
at this temperature, the Parinello-Rahman technique was employed to
perform simulations at constant pressure ($p=0$). After the
determination of the lattice constant, the configuration was
re-equilibrated with a fixed volume. All simulations were carried out
using the velocity Verlet algorithm with a time step $\Delta t =
0.5\,\mathrm{fs}$. In simulations of the bulk configuration, periodic
boundary conditions were applied.

For the calculations of vibrational properties from our simulations we
relied on the velocity-autocorrelation function. The total VDOS
$g(\nu)$ of a system of atoms is proportional to the Fourier-transform
of the velocity-autocorrelation function averaged over all atoms. The
normalized VDOS of an $N$ atom system is thus given by
\begin{equation} g(\nu) = \int_{-\infty}^{\infty} \mathrm{d}t
\frac{\sum_{i=1}^N\left\langle \mathbf{v}_i(t) |
\mathbf{v}_i(0)\right\rangle} {\sum_{i=1}^N\left\langle
\mathbf{v}_i(0) | \mathbf{v}_i(0)\right\rangle} 
\mathrm{e}^{i\,2\pi\nu t}
\end{equation} where $\mathbf{v}_i$ denotes the velocity of the $i$-th
atom. Information about individual vibrational frequencies can be
obtained by this method if the velocities of the atoms are projected
onto a plane wave.

For a mode with wave-vector $\mathbf{q}$ and a normalized polarization
vector $\mathbf{p}$ we calculated the function
\begin{equation} g_\mathbf{q}^p(\nu) = \int_{-\infty}^{\infty}
\mathrm{d}t \frac{\left\langle {v}_\mathbf{q}^p(t) |
{v}_\mathbf{q}^p(0)\right\rangle} {\left\langle {v}_\mathbf{q}^p(0) |
{v}_\mathbf{q}^p(0)\right\rangle} \mathrm{e}^{i\,2\pi\nu t}
\end{equation} where
 \begin{equation} v_\mathbf{q}^p(t) = \sum_{i=1}^{N} \mathbf{p}\cdot
\mathbf{v}_i(t)\,\mathrm{e}^{-i\,\mathbf{q}\cdot\mathbf{r}_i^0}
 \end{equation} is the projection of the velocities on the plane wave
and $\mathbf{r}_i^0$ denotes the average position of the $i$-th
atom. For a crystalline system with translational symmetry the
function $g_\mathbf{q}^p$ should have sharp peaks at the positions of
the phonons with wave-vector $\mathbf{q}$ that have a component along
the polarization vector $\mathbf{p}$.

The vibrational modes of systems without translational symmetry are
not plane waves. For systems similar to perfect crystals (for example
a finite piece cut out of an infinite crystal) one finds however that
the vibrational modes are well described by phonon-like
wave-packets. Such wave-packets appear in $g_\mathbf{q}^p(\nu)$ as
more or less broadened peaks whose maxima can still be used to
determine the dominant frequency of the mode. The quality of the peaks
in $g_\mathbf{q}(\nu)$ --- if there are clear, isolated peaks --- can
therefore be used to judge to what extent the phonon picture applies
to the system in question.

%
%
\section{Results}
%
%
\begin{figure}
\includegraphics[width=8.4cm]{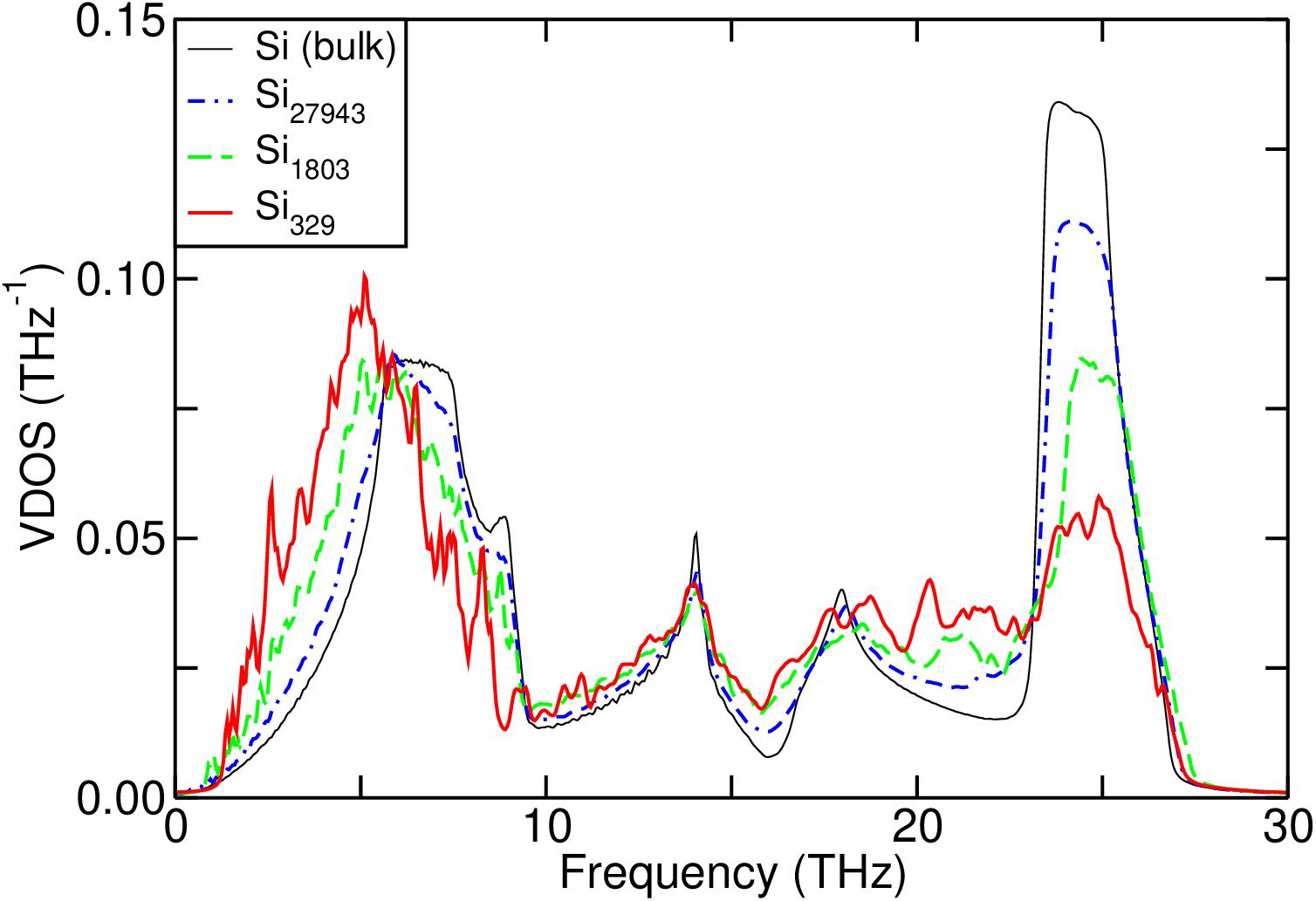}
\caption{(Color online) Normalized vibrational density of states of
crystalline bulk Si and nanoparticles of various sizes. The data have
been broadened with a finite line-width of corresponding to an energy
of $\approx$0.12\,THz}
\label{FigVdosFull}
\end{figure} In Fig.~\ref{FigVdosFull} we compare the VDOS that we
obtained for three selected nanoparticles with the result of
crystalline bulk Si. Since for the smaller particles the discrete
nature of the spectrum becomes notable, we have broadened the VDOS
data with a finite line-width corresponding to an energy of 0.5\,meV
(0.12\,THz). As observed previously in Ref.~\onlinecite{Heino:07a},
the MEAM potential from Ref.~\onlinecite{Baskes:94a} leads
quantitatively to strongly exaggerated phonon
frequencies. Qualitatively however, the results show a clear evolution
of the VDOS as the particle size is diminished with the strongest
changes occurring at the upper and lower end of the frequency
spectrum.

It can be seen from Fig.~\ref{FigVdosFull} that with decreasing
particle size the large peak of transverse acoustic (TA) modes in the
rage 0 -- 10\,Thz is shifted to lower frequencies. The resulting
increase of the VDOS is further amplified by the fact that the curves
become more linear for the smaller particles. These two changes lead
together to a significant increase of the low-frequency VDOS. There is
however no notable loss of the height of the TA mode peak.

%
%
\begin{figure}
\includegraphics[width=8.4cm]{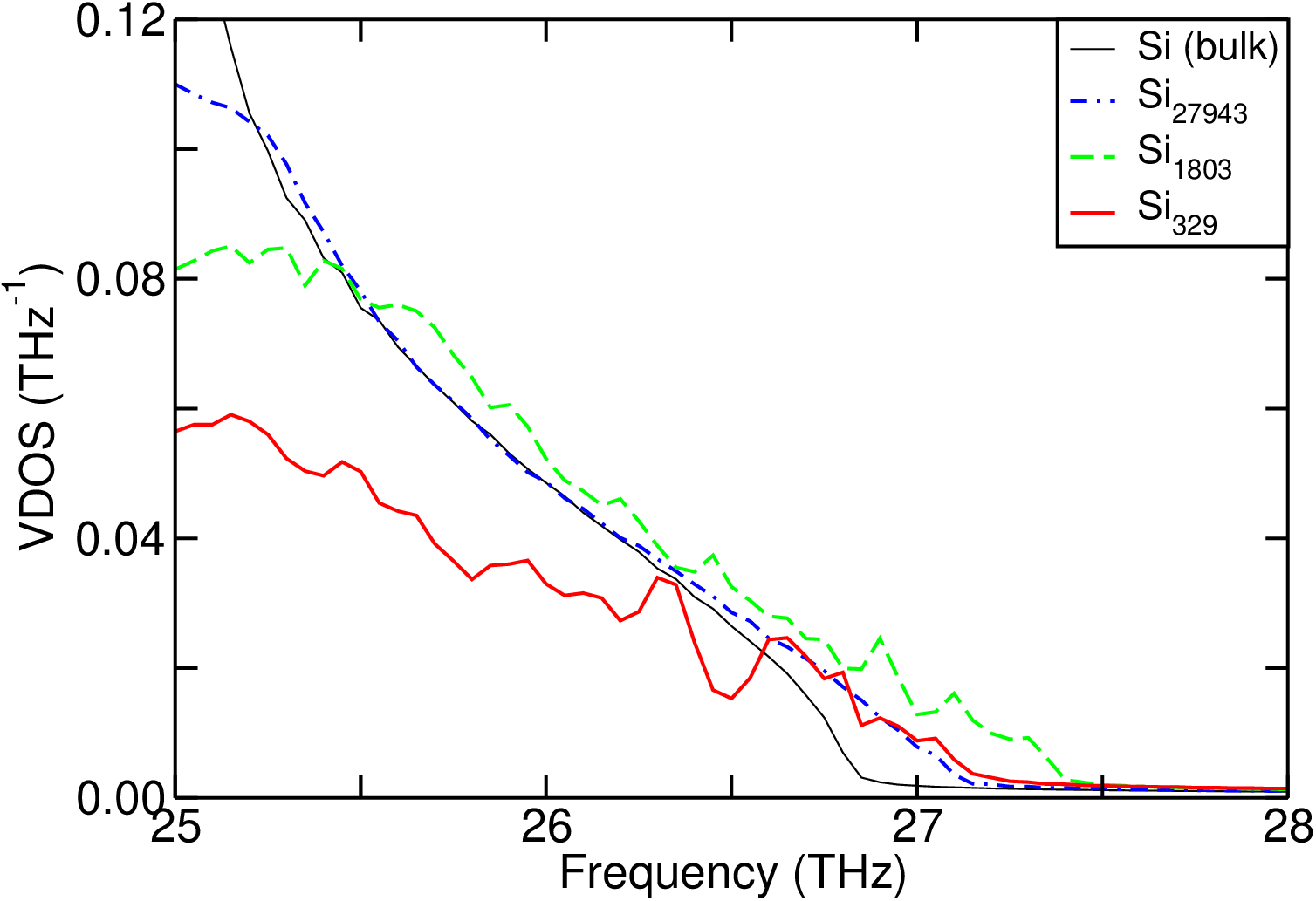}
\caption{(Color online) High frequency part of the normalized
vibrational density of crystalline bulk Si and nanoparticles of
various sizes.}
\label{FigVdosHigh}
\end{figure} In the high-frequency range the VDOS of the Si
nanoparticles show a different behavior. As shown by
Fig.~\ref{FigVdosFull}, the strong peak of the TO modes around 25\,THz
loses its height rapidly with increasing particle size without a
significant increase in peak width. In addition to this, the TO peak
is slightly shifted to higher frequencies. This can also be seen in
Fig.~\ref{FigVdosHigh} where we show the high frequency flank of the
TO peak without line broadening. This figure makes it clear that for
the nanoparticle systems this flank of the peak and the cutoff
frequency are located at higher frequencies than in the bulk
system. Due to the lower peak height the VDOS curves of the particles
eventually cross the bulk line at frequencies between 25 and 26.5\,THz
but it is clear from Fig.~\ref{FigVdosFull} that this is not due to a
peak broadening but a shift to higher frequencies. The shift of the TO
peak will be discussed in more detail below by looking at the
frequencies of individual vibrational modes.

Interestingly we observe a non-monotonous behavior of the cut-off
frequencies as shown by the fact that the cut-off frequencies of the
$\mathrm{Si}_{329}$ and $\mathrm{Si}_{27943}$ particles are both lower
than that of $\mathrm{Si}_{1803}$. We interpret this as an indication
that our smallest particles are below the limit of the scaling regime
in which the physical properties scale monotonously with the particle
size in a simple manner. This view is confirmed by other results
discussed below.

%
%
At intermediate frequencies, the VDOS data presented in
Fig.~\ref{FigVdosFull} show only little differences between the
nanoparticles and the bulk system. In particular, the shape and the
position of the longitudinal acoustic (LA) modes peak at approximately
14\,THz remains remarkably constant for all of our model systems. On
the other hand, the position of the longitudinal optical (LO) mode
peak at 18\,THz is slightly shifted to higher frequencies. In the
regions to the left and to the right of the LO peak the VDOS increases
with decreasing particle sizes. This increase accounts obviously for
the loss of weight in the TO peak.

%
%
\begin{figure}
\includegraphics[width=8.4cm]{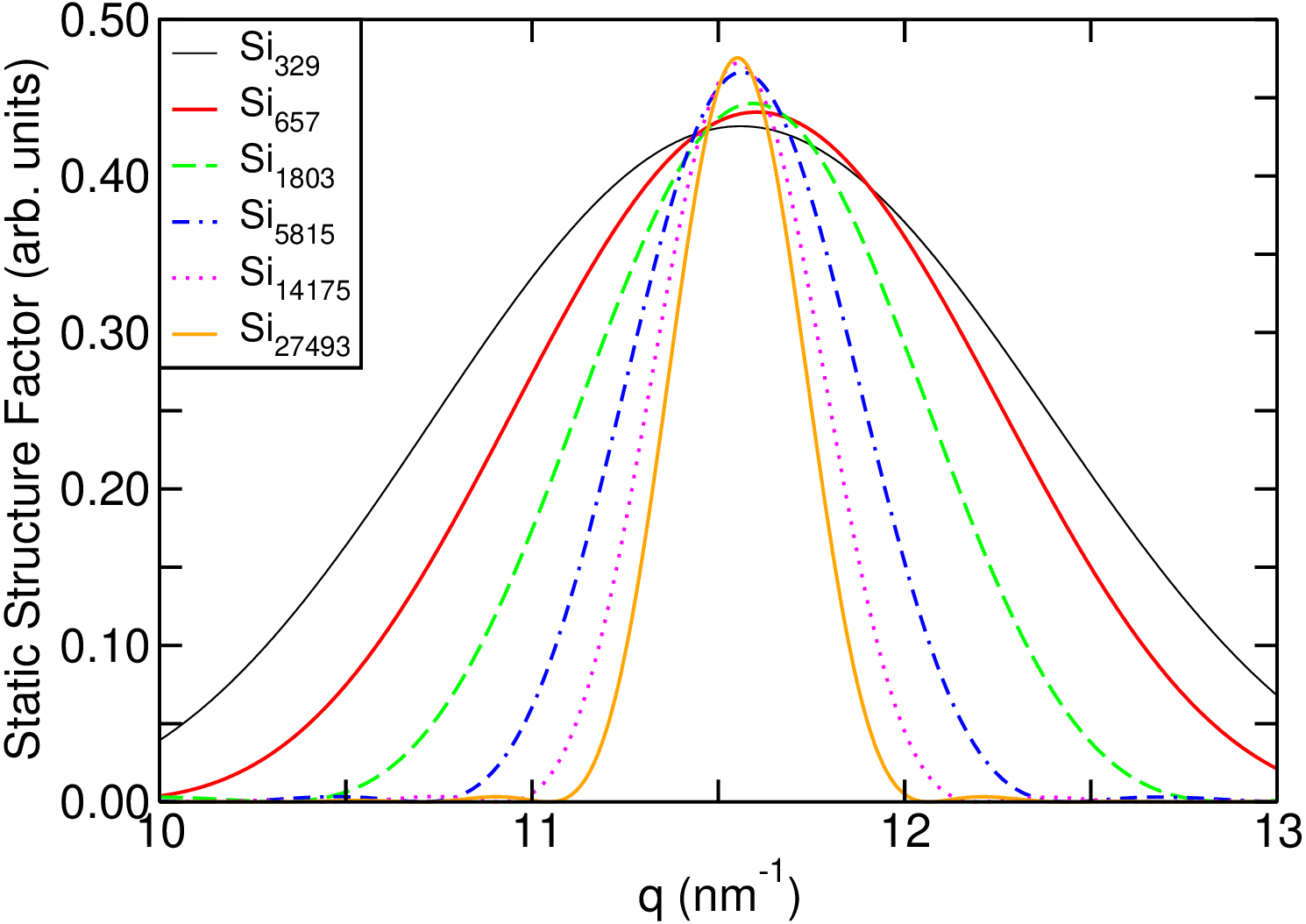}
\caption{(Color online) Static structure factor along the line
$\mathbf{q} = [qqq]$ for Si nanoparticles of various sizes.  }
\label{FigStatic}
\end{figure} Capillary pressure is a pressure in a material that is
caused by curved surfaces or interfaces. The magnitude of this
pressure depends upon the curvature of the surface and the materials
surface or interface stress (which is independent from the
curvature). While this phenomenon is well known and easy to understand
in the case of fluids, the situation is slightly more complex in the
case of crystalline solid materials. A detailed overview of the
effects of curved surfaces in solid systems has been given by Wolf and
Nozi\`{e}re.\cite{Wolf:88a}

In nanoparticles capillary pressure can have a notable effect on the
lattice constant insdie the particle since the small size of the
particles leads to extremely high surface curvatures. Such effects
have previously been observed theoretically an experimentally in
metallic nanoparticles (see for example Ref. \onlinecite{Meyer:03a}
and references therein). In order to account for the effect of
capillary pressure as well as surface relaxation effects on the
lattice constant in our calculations of individual vibrational modes
at high-symmetry points of the Brillouin zone, we have calculated the
static structure factor of the model particles and obtained the
lattice constant from the position of the $(111)$ peak.

Figure~\ref{FigStatic} shows the static structure factor along the
$[111]$ direction for all of our model particles. Except for the
smallest particle, the maximum of the $(111)$ peak shifts to larger
values of $q$ with decreasing particle size. This is the expected
behavior since the smaller particles should have a larger capillary
pressure (assuming a positive surface stress for Si) and consequently
a smaller lattice constant. The fact that the position of the maximum
of the $\mathrm{Si}_{329}$ particle moves against the trend is another
indication that this particle is outside the scaling regime for Si
particles. It should be understood here that the lattice constants
derived from $q_\mathrm{max}$ are only meant to be used as
approximate, average values since the crystal lattice in the
nanoparticles is not homogenous. Capillary pressure compresses the
lattice inside the particles whereas the surface stress has the
inverse effect in the vicinity of the surface.

%
%
\begin{figure}
\includegraphics[width=8.4cm]{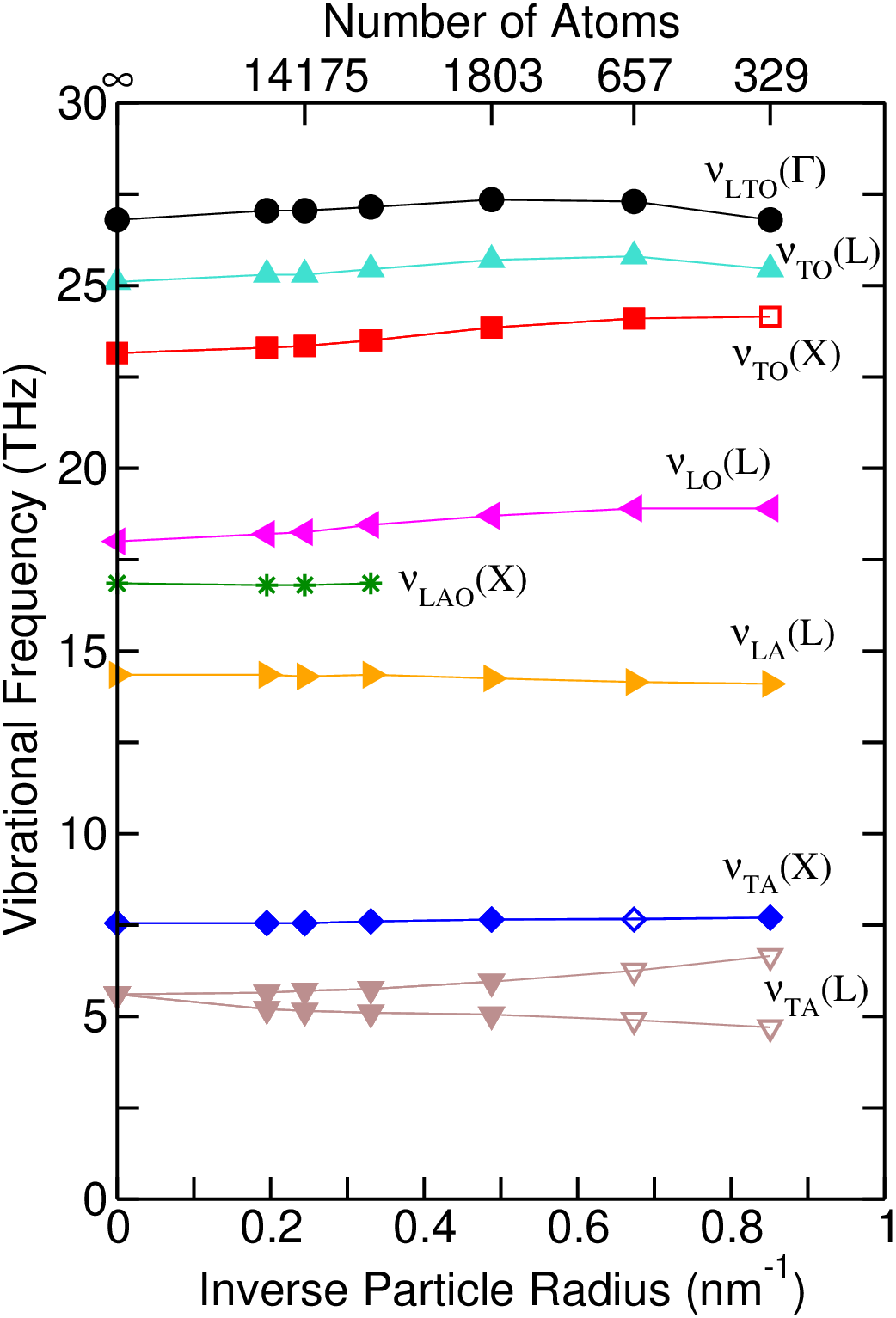}
\caption{(Color online) Frequencies of selected Si nanoparticle phonon
modes as a function of the inverse particle radius. Data at
$1/R=0\,\mathrm{nm}^{-1}$ represent crystalline bulk values. For
further discussions see main article text.  }
\label{FigPhon}
\end{figure} With the average lattice constants derived from the
static structure factor calculations, we calculated the frequencies of
the vibrational modes corresponding to the $\Gamma$, X, and L point of
the Brillouin-zone of the diamond lattice for all of our model
nanoparticles using the method outlined in Sec.~\ref{SecComp}. The
frequencies obtained from the maxima of the peaks in the projected
density of states $g_\mathbf{q}^p(\nu)$ are presented in
Fig.~\ref{FigPhon}. Note that at the $\Gamma$ point the LO and TO
modes and at the X point the LA and LO modes are degenerate. For this
reason the frequencies of these modes are denoted by
$\nu_\mathrm{LTO}(\Gamma)$ and $\nu_\mathrm{LAO}$(X), respectively.

For the larger particles and the bulk system the analysis procedure
gave straightforward, unambiguous results indicating that the
vibrational modes are sufficiently similar to plane waves. In case of
the smaller particles $\mathrm{Si}_{329}$ and $\mathrm{Si}_{657}$,
however, the projected density of states sometimes showed a group of
similar peaks instead of a single peak. If in such a case it was still
possible to identify one peak that was clearly most consistent with
the rest of the data we marked the corresponding data point with an
open symbol in Fig.~\ref{FigPhon}. In the case of the
$\nu_\mathrm{LAO}$(X) mode, however, it turned out to be impossible to
identify matching peaks for all particles smaller than
$\mathrm{Si}_{5815}$ so that these points were dropped from the
figure.

A special case is the $\nu_\mathrm{TA}$(L) mode. For this mode, the
projected density of states of all nanoparticles showed two clearly
distinguishable peaks. This indicates that the degeneracy of this mode
is lifted in the finite particles. For this reason we included two
distinct data sets for this mode in Fig.~\ref{FigPhon}.

The behavior of the individual frequencies is generally consistent
with the changes in the total density of states shown in
Fig.~\ref{FigVdosFull}. In the bulk system, the TA modes at the L and
X point define the upper and lower boundary of the flat plateau on top
of the TA peak. The broadening and shift of this peak in the total
VDOS compares well with the fact that $\nu_\mathrm{TA}$(X) remains
nearly constant, whereas the lower branch of the L-point TA mode
shifts to lower frequencies for smaller particles.

In the intermediate frequency regime, the sharp maxima of the LA and
LO mode peaks correspond to $\nu_\mathrm{LA}$(L) ad
$\nu_\mathrm{LO}$(L) in the bulk case. Again, the very small shift of
$\nu_\mathrm{LA}$(L) and the slightly stronger upward shift of
$\nu_\mathrm{LO}$(L) compare well with the changes in the total
VDOS. This indicates that these modes are well described by the
bulk-phonon picture.

At high frequencies, $\nu_\mathrm{TO}$(X) and $\nu_\mathrm{TO}$(L)
correspond to the lower and upper bound of the TO peak in the bulk
case, whereas $\nu_\mathrm{LTO}(\Gamma)$ defines the cut-off
frequency. The general increase of these frequencies with decreasing
particle size (except for $\mathrm{Si}_{329}$) reflects the shift of
the TO peak and the cut-off point in the same direction as shown in
Fig.~\ref{FigVdosFull} and \ref{FigVdosHigh}. Moreover, the data in
Fig.~\ref{FigPhon} support a slight sharpening of the TO peak; however
the diminishing height of this peak in the VDOS makes it hard to
verify this in Fig.~\ref{FigVdosFull}. The slight decrease of the TO
mode frequencies in the case of the smallest particle gives further
evidence of the growing deviations from the bulk behavior in particles
with diameters below 3\,nm.

Although the general trends of Fig.~\ref{FigPhon} agree well with the
behavior of the VDOS, one has to be careful not to stretch these
comparisons too far. As pointed out in the last paragraph,
$\nu_\mathrm{LTO}(\Gamma)$ represents the cutoff frequency in the bulk
case. In case of $\mathrm{Si}_{329}$, however, the calculated value of
$\nu_\mathrm{LTO}(\Gamma) = 26.80\,\mathrm{THz}$ is the same as in the
bulk case, although Fig.~\ref{FigVdosHigh} clearly shows that modes
with higher frequencies are present in this cluster. This discrepancy
is another hint that this particle is close to the limit of the regime
where the phonon picture can be applied successfully.

%
%
 \begin{figure}
\includegraphics[width=8.4cm]{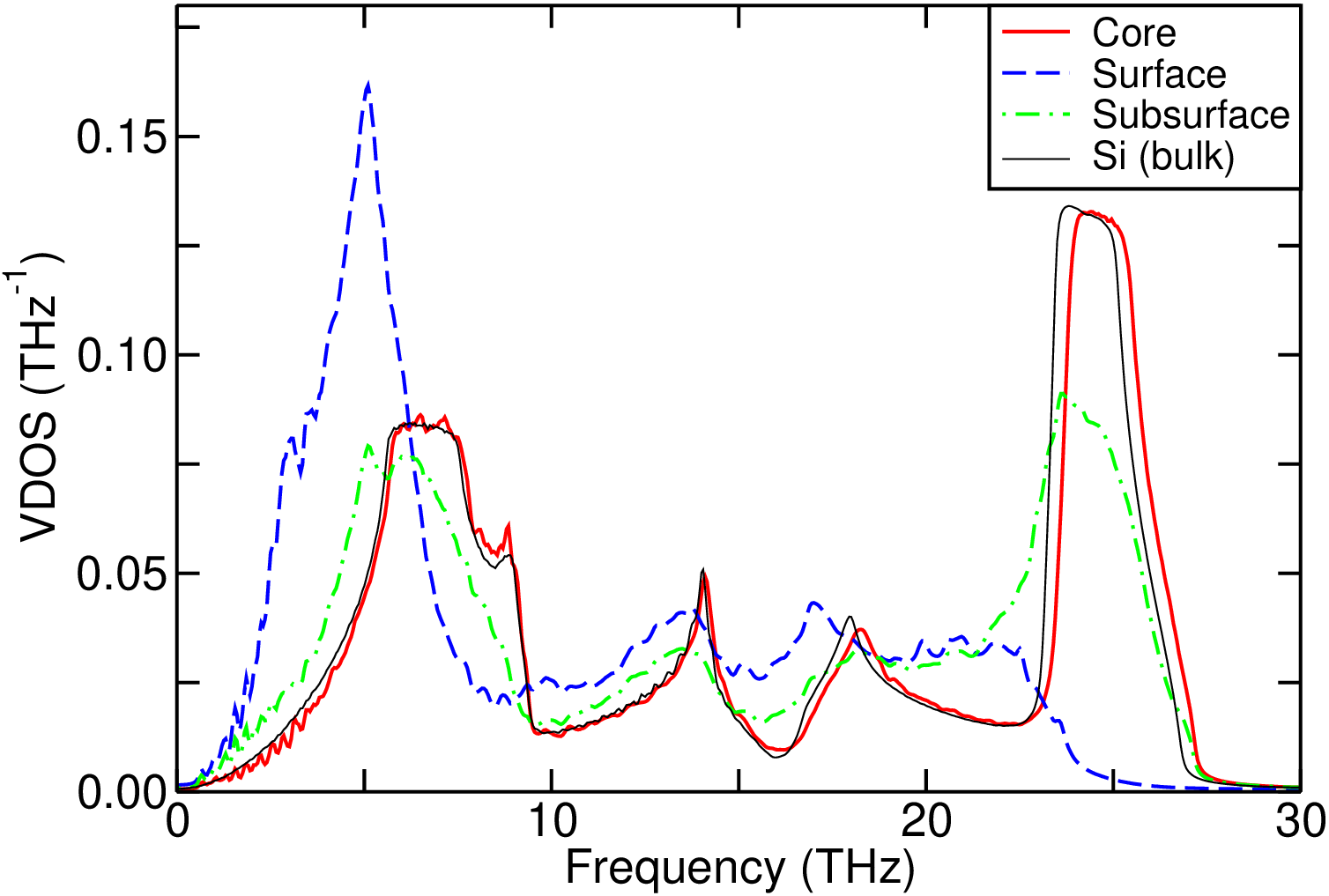}
\caption{(Color online) Normalized local vibrational density of states
in a $\mathrm{Si}_{5815}$ nanoparticle compared to the density of
states of bulk Si.}
\label{FigPDOS}
\end{figure} With decreasing size of a nanoparticles, the fraction of
atoms that is located at or near its surface increases at the expense
of the fraction of atoms in the interior of the particle. This means
that the contribution of surface modes to the total VDOS becomes
increasingly important for the smaller particles.

In order to get insight into the influence of surface modes on the
total VDOS, we have classified the atoms of one of our particles
($\mathrm{Si}_{5815}$) into three different groups: core atoms with a
distance to the center $d < r_\mathrm{c} = 2.35\,$\,nm, surface atoms
with a coordination number $Z<4$ and subsurface atoms with a perfect
coordination $Z=4$ but $d \ge r_\mathrm{c}$. The subsurface layer is
necessary since atoms that are close to the surface generally do not
behave in the same way as atoms deep in the particle. The core,
surface, and subsurface groups contain 2809, 835, and 2171 atoms,
respectively. For each of the three groups we then calculated its
contribution to the total VDOS of the particles. The resulting partial
or local VDOS of the three regions in the $\mathrm{Si}_{5815}$
particle are shown in Fig.~\ref{FigPDOS} together with the VDOS of
bulk Si.

The partial VDOS shown in Fig.~\ref{FigPDOS} reveal substantial
differences between the three groups of atoms. The VDOS of the core
atoms is very similar to the bulk curve. Below 15\,Thz, in the range
of the TA and LA modes, the two curves are nearly identical whereas
the TA and TO peaks of the core atoms are slightly shifted to higher
frequencies. This behavior of the core atoms agrees well with the
trends of the bulk-like modes shown in Fig.~\ref{FigPhon}.

The VDOS of the surface atoms in Fig.~\ref{FigPDOS} is markedly
different from the bulk VDOS. At low frequencies a strong peak can be
seen at frequencies below the TA peak of the bulk and core atoms. At
intermediate frequencies the peak of the surface atoms LO modes
appears to be shifted to lower frequencies and a broad band of modes
is present around 20\,Thz. At high frequencies, in the range of the TO
modes of the bulk and core atoms, however, the surface VDOS is very
low.

A comparison of the trends observed in Fig.~\ref{FigVdosFull} with the
surface VDOS in Fig.~\ref{FigPDOS} shows that in those areas where
the surface VDOS in $\mathrm{Si}_{5815}$ is notably larger than the
bulk VDOS, the total VDOS increases for the smaller particles. We
therefore attribute the shift of the TA peak to lower frequencies, the
increases of the VDOS at intermediate frequencies as well as the
marked decrease of the TO peak to the growing influence of surface
modes on the total VDOS. Not surprisingly, the subsurface atoms show a
behavior somewhere between the two other groups.

%
%
\section{Summary and Conclusions} We have studied the vibrational
properties of silicon nanoparticles with diameters in the range
2.3-10.3\,nm using molecular-dynamics simulations. Our results show
clear changes of the VDOS as the particles become smaller. The
strongest changes are visible in the low- and high-frequency part of
the spectrum. At low frequencies, the VDOS increases as the particles
become smaller and the TA mode peak is shifted to lower energies
whereas at the upper end of the spectrum, the height of the TO mode
peak is significantly reduced although the peak is shifted slightly
($<1\,\mathrm{THz}$) to higher frequencies. Finally, an increase of
the total VDOS is observed at intermediate frequencies between the
peaks of the LA and TO bands.

Our calculations of the frequencies of individual vibrational modes
and partial VDOS show that the observed VDOS changes are driven by two
distinct mechanisms. The first mechanism is the growing fraction of
surface atoms in the smaller particles which increases the
contribution of surface modes to the total VDOS. As shown by the
partial VDOS of the surface atoms in Fig.~\ref{FigPDOS} this effect
accounts for the shift of the LA peak to lower frequencies, the
increase of the VDOS at intermediate frequencies and the notable
decrease of the TO peak.

While the surface modes lead to an overall downward shift of the VDOS,
the picture changes if one looks at the frequency change of specific
bulk-like phonon modes. As shown by Fig.~\ref{FigPhon} and confirmed
by the behavior of the core atoms in Fig.~\ref{FigPDOS} (the partial
VDOS of the core atoms should not be affected by surface modes), the
frequencies of the bulk modes are either unchanged or increase as the
particles become smaller.  The most likely reason for this behavior is
the presence of a capillary pressure inside the core of the particles
that compresses and stiffens the lattice.

In summary one can say that there are two different surface effects
that affect the VDOS of Si nanoparticles in an opposing manner. While
the appearance of surface modes that are not present in bulk Si pushes
the VDOS of the particles to lower frequencies, the frequencies of the
remaining bulk modes are increased by the effect of capillary
pressure.

The increase of the VDOS at intermediate frequencies at the expense of
the height of the TO peak that is caused by the surface modes provides
a good explanation for the experimentally observed strong decrease of
Raman shifts in silicon nanoparticles.\cite{Meier:06a} Calculations of
Meier \textit{et al.}\cite{Meier:06a} have shown that the measured
Raman frequencies are lower than those predicted by the phonon
confinement model. Their calculations, however, used the dispersion
relations of bulk silicon and did not include any changes of the VDOS
in the nanoparticles. Our results suggests that the transfer of weight
from the TO peak to lower frequencies caused by the appearance of
surface modes can account for the discrepancy between the experimental
results and the phonon confinement model.

In general, our calculations agree well with previous studies of the
vibrational properties of fcc and bcc metal
nanoparticles.\cite{Kara:98a,Sun:01a,Meyer:02a,Meyer:03a,Meyer:07a,Roldan:07a}
Similar to the observations in this study, the metallic nanoparticles
show an increase and more linear behavior of the VDOS at low
frequencies whereas the high frequency LA peak (note that there are no
optical modes in the fcc and bcc lattices) is diminished and shifted
to higher frequencies. Partial density of states calculations have
shown that the increase of the low-frequency VDOS and the loss of
weight of the LA peak in copper nanoparticles can be attributed to
modes localized at the particle surfaces\cite{Meyer:03a,Meyer:07a}
whereas the shift of the LA peak is caused by capillary
pressure.\cite{Meyer:02a,Meyer:03a} We conclude that the TA and TO
peaks in the Si nanoparticles undergo similar changes as the TA and LA
peaks in the metallic systems.

The structure of the calculated VDOS and the fact that we were able to
derive phonon frequencies from the projected density of states shows
that the phonon picture can be successfully used to describe the
vibrational modes in Si nanoparticles with diameters down to sizes
below 4\,nm. For particles with diameters of 3\,nm and less our
results indicate an onset of non-monotonous changes of physical
properties as a function of the particle size. This means that the
size of these particles is below the lower limit of the scaling regime
where properties scale with the particle diameter in a simple manner.
The vibrational modes of particles in this range are still rather
bulk-like, but deviations from the bulk picture become more
pronounced. This finding agrees with Ref.~\onlinecite{Valentin:07a}
where it was concluded that for silicon particles with diameters above
2.5\,nm the VDOS becomes bulk-like.

Unfortunately, the limitations of the MEAM potential employed in this
work do not allow calculations of quantitatively correct phonon
frequencies. For this reason it would be interesting to verify our
results with another model. While DFT calculations of the larger
particles are currently out if reach, it might be possible to perform
such calculations for the smallest particles used in this study.

%
%
\begin{acknowledgments} R.M. would like to thank A. Lorke and C. Meier
for fruitful discussions and insight into their experimental
results. We also thank P. Entel for his discussions about our
simulations.  Support by the Shared Hierarchical Academic Research
Network (SHARCNET),\cite{SHARCNET} Laurentian University, and the
\textit{Deutsche Forschungsgemeinschaft} (SFB 445) is gratefully
acknowledged.
\end{acknowledgments}
%
%
%

\begin{thebibliography}{10}%
\makeatletter
\providecommand \@ifxundefined [1]{%
 \ifx #1\undefined \expandafter \@firstoftwo
 \else \expandafter \@secondoftwo
\fi
}%
\providecommand \@ifnum [1]{%
 \ifnum #1\expandafter \@firstoftwo
 \else \expandafter \@secondoftwo
\fi
}%
\providecommand \enquote [1]{``#1''}%
\providecommand \bibnamefont  [1]{#1}%
\providecommand \bibfnamefont [1]{#1}%
\providecommand \citenamefont [1]{#1}%
\providecommand\href[0]{\@sanitize\@href}%
\providecommand\@href[1]{\endgroup\@@startlink{#1}\endgroup\@@href}%
\providecommand\@@href[1]{#1\@@endlink}%
\providecommand \@sanitize [0]{\begingroup\catcode`\&12\catcode`\#12\relax}%
\@ifxundefined \pdfoutput {\@firstoftwo}{%
 \@ifnum{\z@=\pdfoutput}{\@firstoftwo}{\@secondoftwo}%
}{%
 \providecommand\@@startlink[1]{\leavevmode}%
 \providecommand\@@endlink[0]{}%
}{%
 \providecommand\@@startlink[1]{%
  \leavevmode
  \pdfstartlink
   attr{/Border[0 0 1 ]/H/I/C[0 1 1]}%
   user{/Subtype/Link/A<</Type/Action/S/URI/URI(#1)>>}%
  \relax
 }%
 \providecommand\@@endlink[0]{\pdfendlink}%
}%
\providecommand \url  [0]{\begingroup\@sanitize \@url }%
\providecommand \@url [1]{\endgroup\@href {#1}{\urlprefix}}%
\providecommand \urlprefix [0]{URL }%
\providecommand \Eprint[0]{\href }%
\@ifxundefined \urlstyle {%
  \providecommand \doi [1]{doi:\discretionary{}{}{}#1}%
}{%
  \providecommand \doi [0]{doi:\discretionary{}{}{}\begingroup
  \urlstyle{rm}\Url }%
}%
\providecommand \doibase [0]{http://dx.doi.org/}%
\providecommand \Doi[1]{\href{\doibase#1}}%
\providecommand \bibAnnote [3]{%
  \BibitemShut{#1}%
  \begin{quotation}\noindent
    \textsc{Key:}\ #2\\\textsc{Annotation:}\ #3%
  \end{quotation}%
}%
\providecommand \bibAnnoteFile [2]{%
  \IfFileExists{#2}{\bibAnnote {#1} {#2} {\input{#2}}}{}%
}%
\providecommand \typeout [0]{\immediate \write \m@ne }%
\providecommand \selectlanguage [0]{\@gobble}%
\providecommand \bibinfo [0]{\@secondoftwo}%
\providecommand \bibfield [0]{\@secondoftwo}%
\providecommand \translation [1]{[#1]}%
\providecommand \BibitemOpen[0]{}%
\providecommand \bibitemStop [0]{}%
\providecommand \bibitemNoStop [0]{.\EOS\space}%
\providecommand \EOS [0]{\spacefactor3000\relax}%
\providecommand \BibitemShut [1]{\csname bibitem#1\endcsname}%
\bibitem{Dolling:63a}%
  \BibitemOpen
  \bibfield{author}{%
  \bibinfo {author} {\bibfnamefont{G.}~\bibnamefont{Dolling}},\ }%
  in\ \emph{\bibinfo {booktitle} {Inelastic Scattering of Neutrons in Solids
  and Liquids}},\ Vol.~\bibinfo {volume} {II},\ \bibinfo {editor} {edited by\
  \bibinfo {editor} {\bibfnamefont{S.}~\bibnamefont{Ekland}}}\ (\bibinfo
  {publisher} {International Atomic Energy Agency},\ \bibinfo {address}
  {Vienna},\ \bibinfo {year} {1963})%
  \bibAnnoteFile{NoStop}{Dolling:63a}%
\bibitem{Nilsson:72a}%
  \BibitemOpen
  \bibfield{author}{%
  \bibinfo {author} {\bibfnamefont{G.}~\bibnamefont{Nilsson}}\ and\ \bibinfo
  {author} {\bibfnamefont{G.}~\bibnamefont{Nelin}},\ }%
  \bibfield{journal}{%
  \bibinfo {journal} {Phys. Rev. B}\ }%
  \textbf{\bibinfo {volume} {6}},\ \bibinfo {pages} {3777} (\bibinfo {year}
  {1972})%
  \bibAnnoteFile{NoStop}{Nilsson:72a}%
\bibitem{Pandey:73a}%
  \BibitemOpen
  \bibfield{author}{%
  \bibinfo {author} {\bibfnamefont{B.~P.}\ \bibnamefont{Pandey}}\ and\ \bibinfo
  {author} {\bibfnamefont{B.}~\bibnamefont{Dayal}},\ }%
  \bibfield{journal}{%
  \bibinfo {journal} {J. Phys. C: Solid State Phys.}\ }%
  \textbf{\bibinfo {volume} {6}},\ \bibinfo {pages} {2943} (\bibinfo {year}
  {1973})%
  \bibAnnoteFile{NoStop}{Pandey:73a}%
\bibitem{Giannozzi:91a}%
  \BibitemOpen
  \bibfield{author}{%
  \bibinfo {author} {\bibfnamefont{P.}~\bibnamefont{Giannozzi}}, \bibinfo
  {author} {\bibfnamefont{S.}~\bibnamefont{de~Gironcoli}}, \bibinfo {author}
  {\bibfnamefont{P.}~\bibnamefont{Pavone}},\ and\ \bibinfo {author}
  {\bibfnamefont{S.}~\bibnamefont{Baroni}},\ }%
  \bibfield{journal}{%
  \bibinfo {journal} {Phys. Rev. B}\ }%
  \textbf{\bibinfo {volume} {43}},\ \bibinfo {pages} {7231} (\bibinfo {year}
  {1991})%
  \bibAnnoteFile{NoStop}{Giannozzi:91a}%
\bibitem{Wei:94a}%
  \BibitemOpen
  \bibfield{author}{%
  \bibinfo {author} {\bibfnamefont{S.}~\bibnamefont{Wei}}\ and\ \bibinfo
  {author} {\bibfnamefont{M.~Y.}\ \bibnamefont{Chou}},\ }%
  \bibfield{journal}{%
  \bibinfo {journal} {Phys. Rev. B}\ }%
  \textbf{\bibinfo {volume} {50}},\ \bibinfo {pages} {2221} (\bibinfo {year}
  {1994})%
  \bibAnnoteFile{NoStop}{Wei:94a}%
\bibitem{Richter:81a}%
  \BibitemOpen
  \bibfield{author}{%
  \bibinfo {author} {\bibfnamefont{H.}~\bibnamefont{Richter}}, \bibinfo
  {author} {\bibfnamefont{Z.~P.}\ \bibnamefont{Wang}},\ and\ \bibinfo {author}
  {\bibfnamefont{L.}~\bibnamefont{Ley}},\ }%
  \bibfield{journal}{%
  \bibinfo {journal} {Solid State Comm.}\ }%
  \textbf{\bibinfo {volume} {39}},\ \bibinfo {pages} {625} (\bibinfo {year}
  {1981})%
  \bibAnnoteFile{NoStop}{Richter:81a}%
\bibitem{Furukawa:88a}%
  \BibitemOpen
  \bibfield{author}{%
  \bibinfo {author} {\bibfnamefont{S.}~\bibnamefont{Furukawa}}\ and\ \bibinfo
  {author} {\bibfnamefont{T.}~\bibnamefont{Miyasato}},\ }%
  \bibfield{journal}{%
  \bibinfo {journal} {Phys. Rev. B}\ }%
  \textbf{\bibinfo {volume} {38}},\ \bibinfo {pages} {5726} (\bibinfo {year}
  {1988})%
  \bibAnnoteFile{NoStop}{Furukawa:88a}%
\bibitem{Sui:92a}%
  \BibitemOpen
  \bibfield{author}{%
  \bibinfo {author} {\bibfnamefont{Z.}~\bibnamefont{Sui}}, \bibinfo {author}
  {\bibfnamefont{P.~P.}\ \bibnamefont{Leong}}, \bibinfo {author}
  {\bibfnamefont{I.~P.}\ \bibnamefont{Herman}}, \bibinfo {author}
  {\bibfnamefont{G.~S.}\ \bibnamefont{Higashi}},\ and\ \bibinfo {author}
  {\bibfnamefont{H.}~\bibnamefont{Temkin}},\ }%
  \bibfield{journal}{%
  \bibinfo {journal} {Appl. Phys. Lett.}\ }%
  \textbf{\bibinfo {volume} {60}},\ \bibinfo {pages} {2086} (\bibinfo {year}
  {1992})%
  \bibAnnoteFile{NoStop}{Sui:92a}%
\bibitem{Munder:92a}%
  \BibitemOpen
  \bibfield{author}{%
  \bibinfo {author} {\bibfnamefont{H.}~\bibnamefont{M\"{u}nder}}, \bibinfo
  {author} {\bibfnamefont{C.}~\bibnamefont{Andrzejak}}, \bibinfo {author}
  {\bibfnamefont{M.~G.}\ \bibnamefont{Berger}}, \bibinfo {author}
  {\bibfnamefont{U.}~\bibnamefont{Klemradt}}, \bibinfo {author}
  {\bibfnamefont{H.}~\bibnamefont{L\"{u}th}}, \bibinfo {author}
  {\bibfnamefont{R.}~\bibnamefont{Herino}},\ and\ \bibinfo {author}
  {\bibfnamefont{M.}~\bibnamefont{Ligeon}},\ }%
  \bibfield{journal}{%
  \bibinfo {journal} {Thin Solid Films}\ }%
  \textbf{\bibinfo {volume} {227}},\ \bibinfo {pages} {27} (\bibinfo {year}
  {1992})%
  \bibAnnoteFile{NoStop}{Munder:92a}%
\bibitem{Fujii:96a}%
  \BibitemOpen
  \bibfield{author}{%
  \bibinfo {author} {\bibfnamefont{M.}~\bibnamefont{Fujii}}, \bibinfo {author}
  {\bibfnamefont{Y.}~\bibnamefont{Kanzawa}}, \bibinfo {author}
  {\bibfnamefont{S.}~\bibnamefont{Hayashi}},\ and\ \bibinfo {author}
  {\bibfnamefont{K.}~\bibnamefont{Yamamoto}},\ }%
  \bibfield{journal}{%
  \bibinfo {journal} {Phys. Rev. B}\ }%
  \textbf{\bibinfo {volume} {54}},\ \bibinfo {pages} {R8373} (\bibinfo {year}
  {1996})%
  \bibAnnoteFile{NoStop}{Fujii:96a}%
\bibitem{Tanino:96a}%
  \BibitemOpen
  \bibfield{author}{%
  \bibinfo {author} {\bibfnamefont{H.}~\bibnamefont{Tanino}}, \bibinfo {author}
  {\bibfnamefont{A.}~\bibnamefont{Kuprin}}, \bibinfo {author}
  {\bibfnamefont{H.}~\bibnamefont{Deai}},\ and\ \bibinfo {author}
  {\bibfnamefont{N.}~\bibnamefont{Koshida}},\ }%
  \bibfield{journal}{%
  \bibinfo {journal} {Phys. Rev. B}\ }%
  \textbf{\bibinfo {volume} {53}},\ \bibinfo {pages} {1937} (\bibinfo {year}
  {1996})%
  \bibAnnoteFile{NoStop}{Tanino:96a}%
\bibitem{Wang:99a}%
  \BibitemOpen
  \bibfield{author}{%
  \bibinfo {author} {\bibfnamefont{N.}~\bibnamefont{Wang}}, \bibinfo {author}
  {\bibfnamefont{Y.}~\bibnamefont{Tang}}, \bibinfo {author}
  {\bibfnamefont{Y.}~\bibnamefont{Zhang}}, \bibinfo {author}
  {\bibfnamefont{C.}~\bibnamefont{Lee}}, \bibinfo {author}
  {\bibfnamefont{I.}~\bibnamefont{Bello}},\ and\ \bibinfo {author}
  {\bibfnamefont{S.}~\bibnamefont{Lee}},\ }%
  \bibfield{journal}{%
  \bibinfo {journal} {Chem. Phys. Lett.}\ }%
  \textbf{\bibinfo {volume} {299}},\ \bibinfo {pages} {237} (\bibinfo {year}
  {1999})%
  \bibAnnoteFile{NoStop}{Wang:99a}%
\bibitem{Saviot:04a}%
  \BibitemOpen
  \bibfield{author}{%
  \bibinfo {author} {\bibfnamefont{L.}~\bibnamefont{Saviot}}, \bibinfo {author}
  {\bibfnamefont{D.~B.}\ \bibnamefont{Murray}},\ and\ \bibinfo {author}
  {\bibfnamefont{M.~C.}\ \bibnamefont{Marco~de Lucas}},\ }%
  \bibfield{journal}{%
  \bibinfo {journal} {Phys. Rev. B}\ }%
  \textbf{\bibinfo {volume} {69}},\ \bibinfo {pages} {113402} (\bibinfo {year}
  {2004})%
  \bibAnnoteFile{NoStop}{Saviot:04a}%
\bibitem{Adu:05a}%
  \BibitemOpen
  \bibfield{author}{%
  \bibinfo {author} {\bibfnamefont{K.~W.}\ \bibnamefont{Adu}}, \bibinfo
  {author} {\bibfnamefont{H.~R.}\ \bibnamefont{Guti\'{e}rrez}}, \bibinfo
  {author} {\bibfnamefont{U.~J.}\ \bibnamefont{Kim}}, \bibinfo {author}
  {\bibfnamefont{G.~U.}\ \bibnamefont{Sumanasekera}},\ and\ \bibinfo {author}
  {\bibfnamefont{P.~C.}\ \bibnamefont{Eklund}},\ }%
  \bibfield{journal}{%
  \bibinfo {journal} {Nano Lett.}\ }%
  \textbf{\bibinfo {volume} {5}},\ \bibinfo {pages} {409} (\bibinfo {year}
  {2005})%
  \bibAnnoteFile{NoStop}{Adu:05a}%
\bibitem{Meier:06a}%
  \BibitemOpen
  \bibfield{author}{%
  \bibinfo {author} {\bibfnamefont{C.}~\bibnamefont{Meier}}, \bibinfo {author}
  {\bibfnamefont{S.}~\bibnamefont{L\"{u}ttjohann}}, \bibinfo {author}
  {\bibfnamefont{V.}~\bibnamefont{Kravets}}, \bibinfo {author}
  {\bibfnamefont{H.}~\bibnamefont{Nienhaus}}, \bibinfo {author}
  {\bibfnamefont{A.}~\bibnamefont{Lorke}},\ and\ \bibinfo {author}
  {\bibfnamefont{H.}~\bibnamefont{Wiggers}},\ }%
  \bibfield{journal}{%
  \bibinfo {journal} {Physica E}\ }%
  \textbf{\bibinfo {volume} {32}},\ \bibinfo {pages} {155} (\bibinfo {year}
  {2006})%
  \bibAnnoteFile{NoStop}{Meier:06a}%
\bibitem{Ristic:09a}%
  \BibitemOpen
  \bibfield{author}{%
  \bibinfo {author} {\bibfnamefont{D.}~\bibnamefont{Risti\'{c}}}, \bibinfo
  {author} {\bibfnamefont{M.}~\bibnamefont{Ivanda}},\ and\ \bibinfo {author}
  {\bibfnamefont{K.}~\bibnamefont{Furi\'{c}}},\ }%
  \bibfield{journal}{%
  \bibinfo {journal} {J. Mol. Struct.}\ }%
  \textbf{\bibinfo {volume} {924--926}},\ \bibinfo {pages} {291} (\bibinfo
  {year} {2009})%
  \bibAnnoteFile{NoStop}{Ristic:09a}%
\bibitem{Campbell:86a}%
  \BibitemOpen
  \bibfield{author}{%
  \bibinfo {author} {\bibfnamefont{I.~H.}\ \bibnamefont{Campbell}}\ and\
  \bibinfo {author} {\bibfnamefont{P.~M.}\ \bibnamefont{Fauchet}},\ }%
  \bibfield{journal}{%
  \bibinfo {journal} {Solid State Comm.}\ }%
  \textbf{\bibinfo {volume} {58}},\ \bibinfo {pages} {739} (\bibinfo {year}
  {1986})%
  \bibAnnoteFile{NoStop}{Campbell:86a}%
\bibitem{Heino:07a}%
  \BibitemOpen
  \bibfield{author}{%
  \bibinfo {author} {\bibfnamefont{P.}~\bibnamefont{Heino}},\ }%
  \bibfield{journal}{%
  \bibinfo {journal} {Eur. Phys. J. B}\ }%
  \textbf{\bibinfo {volume} {60}},\ \bibinfo {pages} {171} (\bibinfo {year}
  {2007})%
  \bibAnnoteFile{NoStop}{Heino:07a}%
\bibitem{Hu:01a}%
  \BibitemOpen
  \bibfield{author}{%
  \bibinfo {author} {\bibfnamefont{X.}~\bibnamefont{Hu}}, \bibinfo {author}
  {\bibfnamefont{G.}~\bibnamefont{Wang}}, \bibinfo {author}
  {\bibfnamefont{W.}~\bibnamefont{Wu}}, \bibinfo {author}
  {\bibfnamefont{P.}~\bibnamefont{Jiang}},\ and\ \bibinfo {author}
  {\bibfnamefont{J.}~\bibnamefont{Zi}},\ }%
  \bibfield{journal}{%
  \bibinfo {journal} {J. Phys.: Condens. Matter}\ }%
  \textbf{\bibinfo {volume} {13}},\ \bibinfo {pages} {L835} (\bibinfo {year}
  {2001})%
  \bibAnnoteFile{NoStop}{Hu:01a}%
\bibitem{Valentin:07a}%
  \BibitemOpen
  \bibfield{author}{%
  \bibinfo {author} {\bibfnamefont{A.}~\bibnamefont{Valentin}}, \bibinfo
  {author} {\bibfnamefont{J.}~\bibnamefont{S\'{e}e}}, \bibinfo {author}
  {\bibfnamefont{S.}~\bibnamefont{Galdin-Retailleau}},\ and\ \bibinfo {author}
  {\bibfnamefont{P.}~\bibnamefont{Dollfus}},\ }%
  \bibfield{journal}{%
  \bibinfo {journal} {J. Phys.: Conf. Ser.}\ }%
  \textbf{\bibinfo {volume} {92}},\ \bibinfo {pages} {012048} (\bibinfo {year}
  {2007})%
  \bibAnnoteFile{NoStop}{Valentin:07a}%
\bibitem{Baskes:89a}%
  \BibitemOpen
  \bibfield{author}{%
  \bibinfo {author} {\bibfnamefont{M.~I.}\ \bibnamefont{Baskes}}, \bibinfo
  {author} {\bibfnamefont{J.~S.}\ \bibnamefont{Nelson}},\ and\ \bibinfo
  {author} {\bibfnamefont{A.~F.}\ \bibnamefont{Wright}},\ }%
  \bibfield{journal}{%
  \bibinfo {journal} {Phys. Rev. B}\ }%
  \textbf{\bibinfo {volume} {40}},\ \bibinfo {pages} {6085} (\bibinfo {year}
  {1989})%
  \bibAnnoteFile{NoStop}{Baskes:89a}%
\bibitem{Baskes:92a}%
  \BibitemOpen
  \bibfield{author}{%
  \bibinfo {author} {\bibfnamefont{M.~I.}\ \bibnamefont{Baskes}},\ }%
  \bibfield{journal}{%
  \bibinfo {journal} {Phys. Rev. B}\ }%
  \textbf{\bibinfo {volume} {46}},\ \bibinfo {pages} {2727} (\bibinfo {year}
  {1992})%
  \bibAnnoteFile{NoStop}{Baskes:92a}%
\bibitem{Baskes:94a}%
  \BibitemOpen
  \bibfield{author}{%
  \bibinfo {author} {\bibfnamefont{M.~I.}\ \bibnamefont{Baskes}}, \bibinfo
  {author} {\bibfnamefont{J.~E.}\ \bibnamefont{Angelo}},\ and\ \bibinfo
  {author} {\bibfnamefont{C.~L.}\ \bibnamefont{Bisson}},\ }%
  \bibfield{journal}{%
  \bibinfo {journal} {Modelling Simul. Mater. Sci. Eng.}\ }%
  \textbf{\bibinfo {volume} {2}},\ \bibinfo {pages} {505} (\bibinfo {year}
  {1994})%
  \bibAnnoteFile{NoStop}{Baskes:94a}%
\bibitem{Kara:98a}%
  \BibitemOpen
  \bibfield{author}{%
  \bibinfo {author} {\bibfnamefont{A.}~\bibnamefont{Kara}}\ and\ \bibinfo
  {author} {\bibfnamefont{T.~S.}\ \bibnamefont{Rahman}},\ }%
  \bibfield{journal}{%
  \bibinfo {journal} {Phys. Rev. Lett.}\ }%
  \textbf{\bibinfo {volume} {81}},\ \bibinfo {pages} {1453} (\bibinfo {year}
  {1998})%
  \bibAnnoteFile{NoStop}{Kara:98a}%
\bibitem{Meyer:02a}%
  \BibitemOpen
  \bibfield{author}{%
  \bibinfo {author} {\bibfnamefont{R.}~\bibnamefont{Meyer}}, \bibinfo {author}
  {\bibfnamefont{S.}~\bibnamefont{Prakash}},\ and\ \bibinfo {author}
  {\bibfnamefont{P.}~\bibnamefont{Entel}},\ }%
  \bibfield{journal}{%
  \bibinfo {journal} {Phase Transitions}\ }%
  \textbf{\bibinfo {volume} {75}},\ \bibinfo {pages} {51} (\bibinfo {year}
  {2002})%
  \bibAnnoteFile{NoStop}{Meyer:02a}%
\bibitem{Meyer:03a}%
  \BibitemOpen
  \bibfield{author}{%
  \bibinfo {author} {\bibfnamefont{R.}~\bibnamefont{Meyer}}, \bibinfo {author}
  {\bibfnamefont{L.~J.}\ \bibnamefont{Lewis}}, \bibinfo {author}
  {\bibfnamefont{S.}~\bibnamefont{Prakash}},\ and\ \bibinfo {author}
  {\bibfnamefont{P.}~\bibnamefont{Entel}},\ }%
  \bibfield{journal}{%
  \bibinfo {journal} {Phys. Rev. B}\ }%
  \textbf{\bibinfo {volume} {68}},\ \bibinfo {pages} {104303} (\bibinfo {year}
  {2003})%
  \bibAnnoteFile{NoStop}{Meyer:03a}%
\bibitem{Meyer:07a}%
  \BibitemOpen
  \bibfield{author}{%
  \bibinfo {author} {\bibfnamefont{R.}~\bibnamefont{Meyer}}\ and\ \bibinfo
  {author} {\bibfnamefont{P.}~\bibnamefont{Entel}},\ }%
  \bibfield{journal}{%
  \bibinfo {journal} {Z. Kristallogr.}\ }%
  \textbf{\bibinfo {volume} {222}},\ \bibinfo {pages} {646} (\bibinfo {year}
  {2007})%
  \bibAnnoteFile{NoStop}{Meyer:07a}%
\bibitem{Baskes:99a}%
  \BibitemOpen
  \bibfield{author}{%
  \bibinfo {author} {\bibfnamefont{M.~I.}\ \bibnamefont{Baskes}},\ }%
  \bibfield{journal}{%
  \bibinfo {journal} {Mater. Sci. Eng.}\ }%
  \textbf{\bibinfo {volume} {A261}},\ \bibinfo {pages} {165} (\bibinfo {year}
  {1999})%
  \bibAnnoteFile{NoStop}{Baskes:99a}%
\bibitem{Kuo:04a}%
  \BibitemOpen
  \bibfield{author}{%
  \bibinfo {author} {\bibfnamefont{C.-L.}\ \bibnamefont{Kuo}}\ and\ \bibinfo
  {author} {\bibfnamefont{P.}~\bibnamefont{Clancy}},\ }%
  \bibfield{journal}{%
  \bibinfo {journal} {Surface Sciences}\ }%
  \textbf{\bibinfo {volume} {551}},\ \bibinfo {pages} {39} (\bibinfo {year}
  {2004})%
  \bibAnnoteFile{NoStop}{Kuo:04a}%
\bibitem{Wolf:88a}%
  \BibitemOpen
  \bibfield{author}{%
  \bibinfo {author} {\bibfnamefont{D.~E.}\ \bibnamefont{Wolf}}\ and\ \bibinfo
  {author} {\bibfnamefont{P.}~\bibnamefont{Nozi\`{e}re}},\ }%
  \bibfield{journal}{%
  \bibinfo {journal} {Z. Phys. B}\ }%
  \textbf{\bibinfo {volume} {70}},\ \bibinfo {pages} {507} (\bibinfo {year}
  {1988})%
  \bibAnnoteFile{NoStop}{Wolf:88a}%
\bibitem{Sun:01a}%
  \BibitemOpen
  \bibfield{author}{%
  \bibinfo {author} {\bibfnamefont{D.~Y.}\ \bibnamefont{Sun}}, \bibinfo
  {author} {\bibfnamefont{X.~G.}\ \bibnamefont{Gong}},\ and\ \bibinfo {author}
  {\bibfnamefont{X.-Q.}\ \bibnamefont{Wang}},\ }%
  \bibfield{journal}{%
  \bibinfo {journal} {Phys. Rev. B}\ }%
  \textbf{\bibinfo {volume} {63}},\ \bibinfo {pages} {193412} (\bibinfo {year}
  {2001})%
  \bibAnnoteFile{NoStop}{Sun:01a}%
\bibitem{Roldan:07a}%
  \BibitemOpen
  \bibfield{author}{%
  \bibinfo {author} {\bibfnamefont{B.}~\bibnamefont{Roldan~Cuenya}}, \bibinfo
  {author} {\bibfnamefont{A.}~\bibnamefont{Naitabdi}}, \bibinfo {author}
  {\bibfnamefont{J.}~\bibnamefont{Croy}}, \bibinfo {author}
  {\bibfnamefont{W.}~\bibnamefont{Sturhahn}}, \bibinfo {author}
  {\bibfnamefont{J.~Y.}\ \bibnamefont{Zhao}}, \bibinfo {author}
  {\bibfnamefont{E.~E.}\ \bibnamefont{Alp}}, \bibinfo {author}
  {\bibfnamefont{R.}~\bibnamefont{Meyer}}, \bibinfo {author}
  {\bibfnamefont{D.}~\bibnamefont{Sudfeld}}, \bibinfo {author}
  {\bibfnamefont{E.}~\bibnamefont{Schuster}},\ and\ \bibinfo {author}
  {\bibfnamefont{W.}~\bibnamefont{Keune}},\ }%
  \bibfield{journal}{%
  \bibinfo {journal} {Phys. Rev. B}\ }%
  \textbf{\bibinfo {volume} {76}},\ \bibinfo {pages} {195422} (\bibinfo {year}
  {2007})%
  \bibAnnoteFile{NoStop}{Roldan:07a}%
\bibitem{SHARCNET}%
  \BibitemOpen
  \url{http://www.sharcnet.ca}%
  \bibAnnoteFile{NoStop}{SHARCNET}%
\end{thebibliography}
\end{document}